\documentclass{aa}
\usepackage{graphics,epsfig,txfonts}
\usepackage{longtable,lscape}
\usepackage{natbib}
\bibpunct{(}{)}{;}{a}{}{,}

\newcommand{\nh}{N$_{\rm H}$}

\newcommand{\HII}{\ion{H}{ii}}
\newcommand{\HI}{\ion{H}{i}}

\newcommand{\ltsima}{$\buildrel < \over \sim$}
\newcommand{\lsim}{\lower.5ex\hbox{\ltsima}}
\newcommand{\gtsima}{$\buildrel > \over \sim$}
\newcommand{\gsim}{\lower.5ex\hbox{\gtsima}}
\newcommand{\xmm}{XMM-{\it Newton}}

\usepackage{color}

\begin{document}

\title{AGN behind the SMC selected from radio and X-ray surveys
       \thanks{Based on observations obtained with XMM-Newton, an ESA science mission with instruments and contributions directly funded by ESA Member States and NASA}
      }

\author{         R.~Sturm\inst{1} 
      \and       D.~Dra{\v{s}}kovi{\'c}\inst{2}
      \and       M.~D.~Filipovi{\'c}\inst{2} 
      \and       F.~Haberl\inst{1}  
      \and       J.~Collier\inst{2}
      \and       E.~J.~Crawford\inst{2}
      \and       M.~Ehle\inst{3} 
      \and       A.~De~Horta\inst{2}
      \and       W.~Pietsch\inst{1}  
      \and       N.~F.~H.~Tothill\inst{2}
      \and       G.~Wong\inst{2}
       }

\titlerunning{AGN behind the SMC selected from radio and X-ray surveys.}
\authorrunning{Sturm et al.}

\institute{ Max-Planck-Institut f\"ur extraterrestrische Physik, Giessenbachstra{\ss}e, 85748 Garching, Germany
	    \and
            University of Western Sydney, Locked Bag 1797, Penrith South DC, NSW1797, Australia
	    \and
            XMM-Newton Science Operations Centre, ESAC, ESA, PO Box 78, 28691 Villanueva de la Ca\~{n}ada, Madrid, Spain
	   }

\date{Received 15 October 2012 / Accepted 20 July 2013}

 \abstract{The \xmm\ survey of the Small Magellanic Cloud (SMC) revealed 3053 X-ray sources with the majority expected to be active galactic nuclei (AGN) behind the SMC. 
           However, the high stellar density in this field often does not allow assigning unique optical counterparts and hinders source classification.
           On the other hand, the association of X-ray point sources with radio emission can be used to select background AGN with high confidence, and to constrain other object classes like pulsar wind nebula.}
          {To classify X-ray and radio sources, we use clear correlations of X-ray sources found in the \xmm\ survey with radio-continuum sources detected with ATCA and MOST.}
          {Deep radio-continuum images were searched for correlations with X-ray sources of the \xmm\ SMC-survey point-source catalogue as well as galaxy clusters seen with extended X-ray emission.}
          {Eighty eight discrete radio sources were found in common with the X-ray point-source catalogue in addition to six correlations with extended X-ray sources. 
            One source is identified as a Galactic star and eight as galaxies.
            Eight radio sources likely originate in AGN that are associated with clusters of galaxies seen in X-rays. One source is a PWN candidate.
            We obtain 43 new candidates for background sources located behind the SMC.
            A total of 24 X-ray sources show jet-like radio structures.}
          {}


\maketitle

\section{Introduction}
\label{sec:introduction}

Searching for background sources in fields with high stellar density like the Small Magellanic Cloud (SMC) can be intricate. 
Once background sources behind the SMC are identified, they constitute a valuable sample of sources. 
Besides studying these sources \citep[e.g.][]{2009ApJ...698..895K}, when multi-wavelength data from several epochs are available, 
they provide an ideal reference frame for astrometry as soon as their positions are known precisely. 
This is important for proper-motion studies of the SMC \citep[e.g.][]{2008AJ....135.1024P}, 
but also to reduce systematic uncertainties in the position of X-ray sources \citep[e.g.][]{2009A&A...493..339W}. 
Further, the interstellar medium of the SMC may be studied with the help of absorption lines in the spectra of illuminators in the background.

The first two quasars behind the SMC were reported by \citet{1982MNRAS.200.1007M} and \citet{1983PASAu...5....2W}.
Later on, \citet{1997MNRAS.285..111T} used optical spectroscopy to confirm additional eight candidates, selected from ROSAT X-ray sources.
\citet{2003AJ....125.1330D,2003AJ....126..734D} added five X-ray selected candidates and five candidates chosen from their optical variability by \citet{2002AcA....52..241E}.
\citet{2009ApJ...701..508K} selected 657 quasar candidates using Spitzer infrared and near-infrared photometry. 
Including also candidates selected from optical variability, \citet{2011ApJS..194...22K,2013arXiv1305.6927K} were able to confirm 193 of 766 observed candidates with followup spectroscopy, 
raising the number of confirmed background quasars to $\sim$200.

In this study, we search for sources with common X-ray and radio emission and classify them. 
The \xmm\ survey of the SMC \citep{2012A&A...545A.128H}, provides for the first time a complete coverage of the SMC main body 
with imaging X-ray optics up to photon energies of 12~keV and with a source-detection sensitivity of $\sim$$2\times 10^{-14}$ erg s$^{-1}$ cm$^{-2}$. 
Compared to previous surveys with ROSAT in the (0.1$-$2.0)~keV band \citep{2000A&AS..142...41H,2000A&AS..147...75S}, 
the sensitivity of \xmm\ at harder X-rays results in the detection of more background sources. 
The higher position accuracy allows a more unique correlation with radio counterparts. 
To identify X-ray and radio sources, we compare our X-ray point-source catalogue with deep merged Australia Telescope radio images of the SMC, 
having unprecedented sensitivity compared to earlier studies \citep[e.g. ][]{1997A&AS..121..321F,1998A&AS..130..421F,2002MNRAS.335.1085F,2004MNRAS.355...44P}. 
Except for a few Galactic stars (such as young stellar objects (YSO) or binary stars) and rare SMC objects like pulsar wind nebulae (PWNe), the bulk of 
discrete sources emitting radio and X-rays are expected to originate in background objects. 
Active galactic nuclei (AGN) produce hard X-ray emission and relativistic jets visible in radio.
In some cases emission can originate in nearby normal galaxies, with little or no contribution of an AGN.
Also the AGN host galaxy can be part of a cluster of galaxies (ClG), where X-rays originate in the hot intracluster medium and radio-continuum emission from the AGN. 
Supernova remnants (SNRs) in the SMC also can show radio and X-ray emission. 
These sources have a significant extent at the distance of the SMC (10\arcsec\ translates to $\sim$3 pc) and can easily be excluded. 
They are not part of this study as these sources will be reviewed in a subsequent paper. 
Other X-ray and radio emitting sources like planetary nebulae are unlikely to be detected at SMC distance \citep[][]{2004MNRAS.355...44P}.

\begin{figure*}
  \resizebox{\hsize}{!}{\includegraphics[angle=-90,clip=]{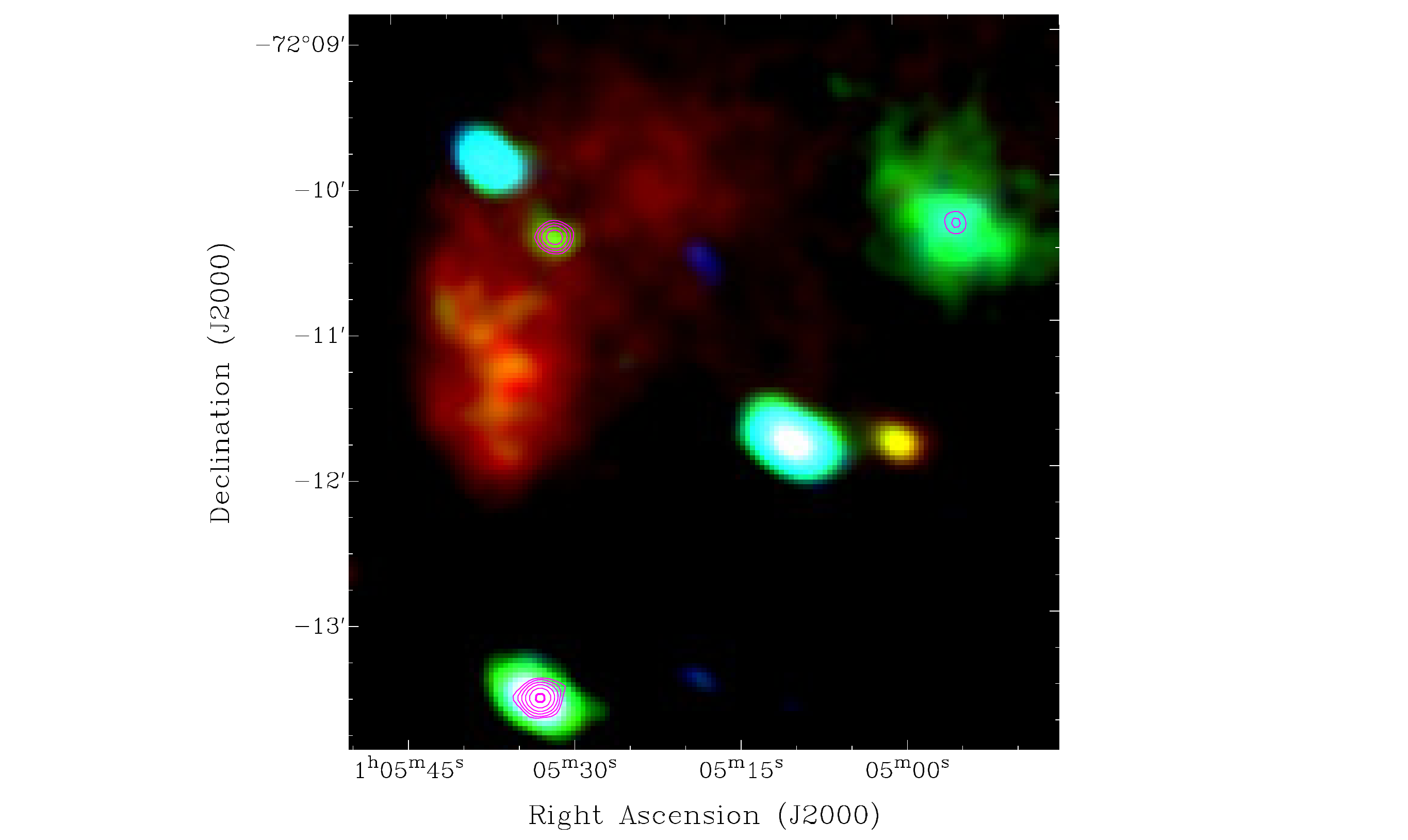}
                        \includegraphics[angle=-90,clip=]{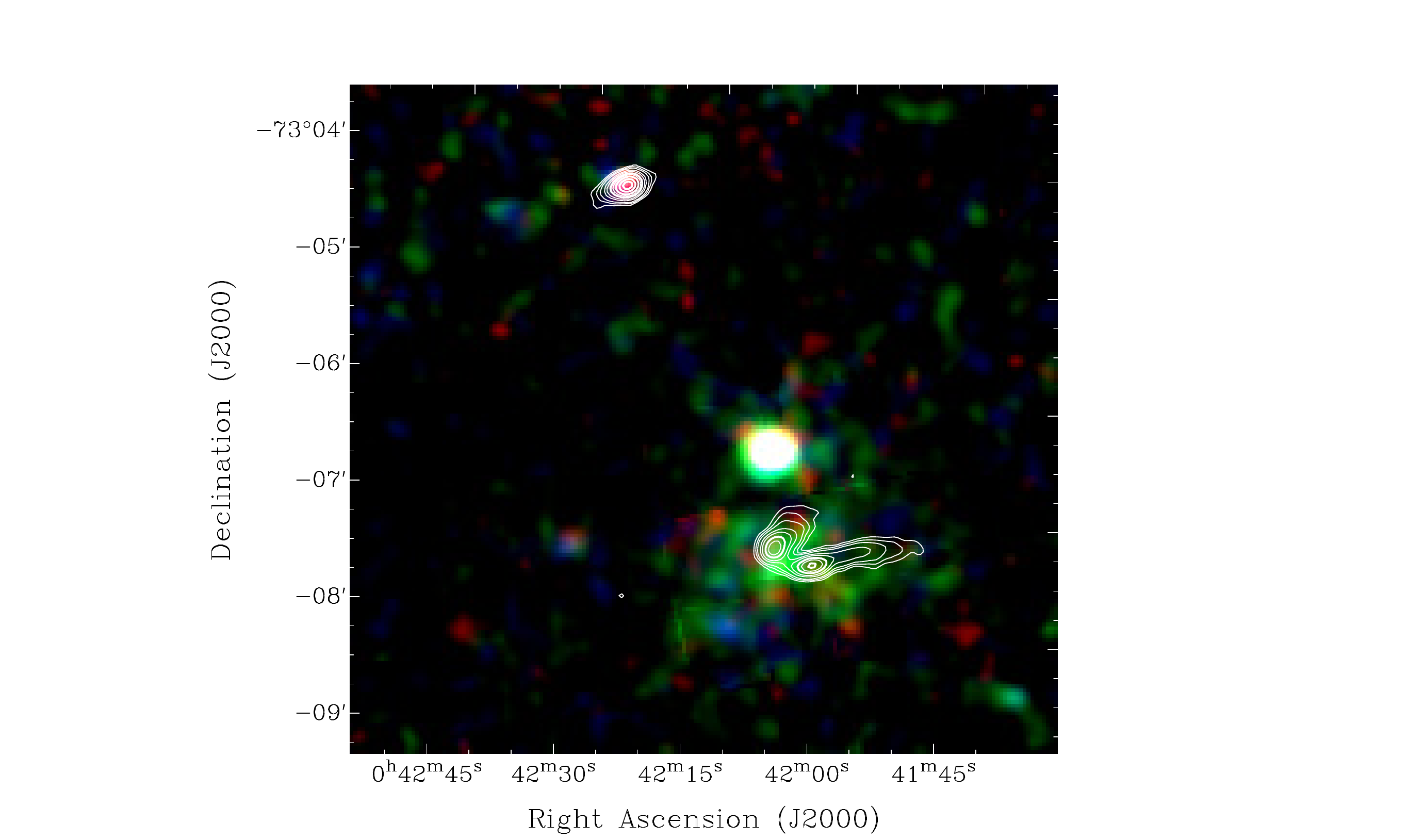}
                       }\\
  \resizebox{\hsize}{!}{\includegraphics[angle=-90,clip=]{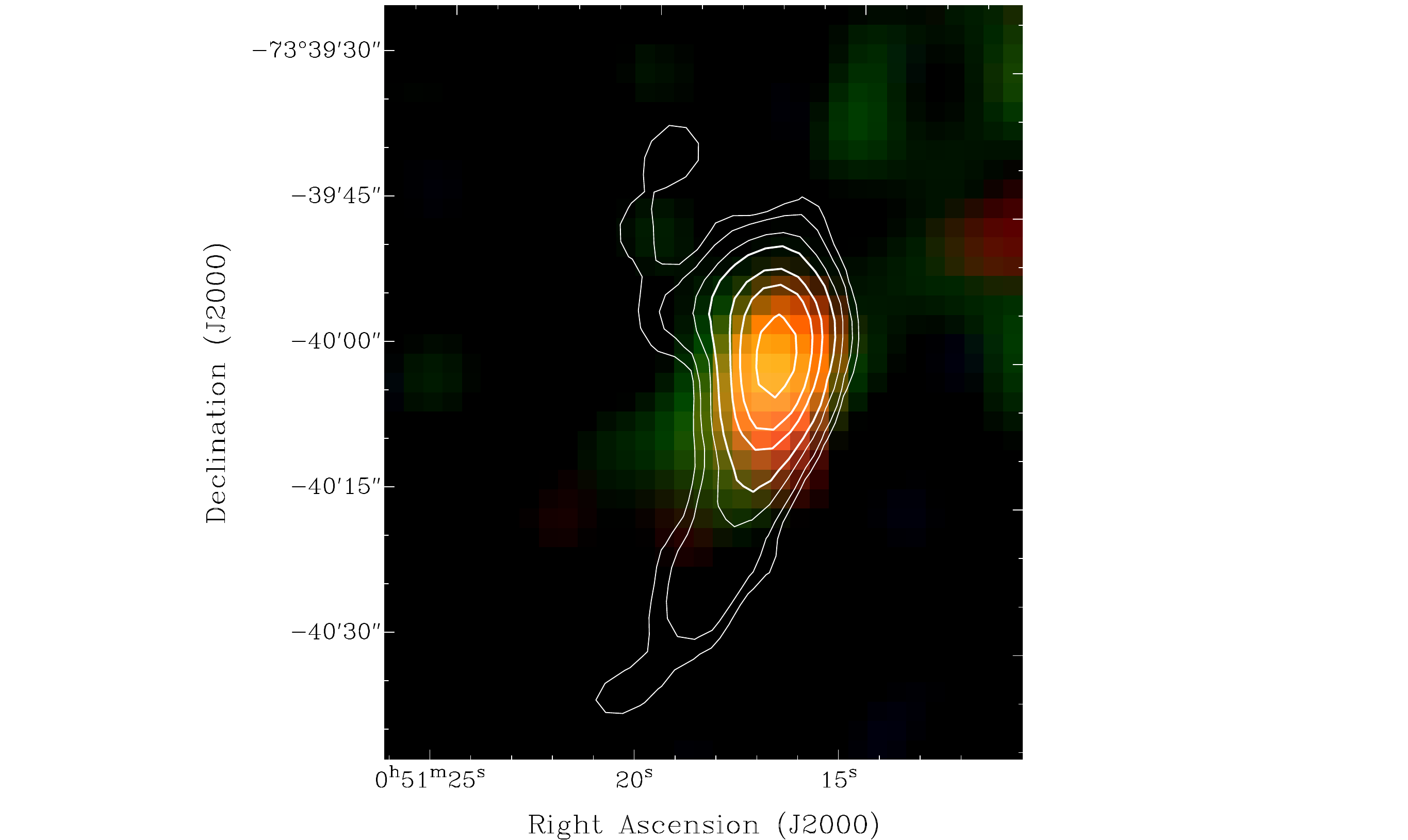}
                        \includegraphics[angle=-90,clip=]{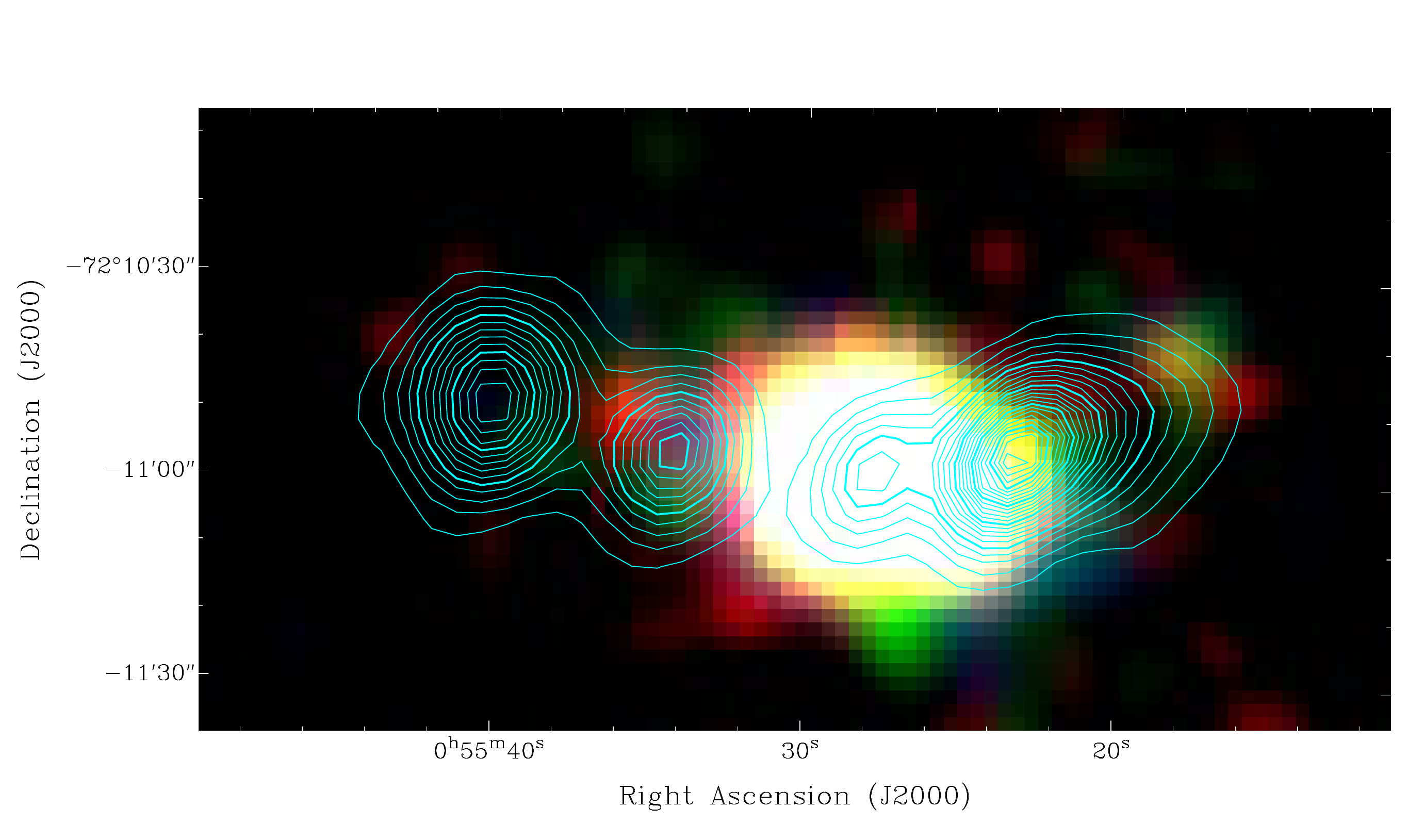}
                       }
  \caption{ EPIC X-ray colour images with overlaid radio contours. Red/green/blue corresponds to intensities in the (0.2--1.0)~keV, (1.0--2.0)~keV and (2.0--4.5)~keV band, respectively. 
           {\it Upper left:}  Three radio sources in the north-east of the SMC are presented in this image. Source No 14 is seen along SNR DEM\,S128 (the red structure). 
                              In the lower left of the image, source No 8 demonstrates an X-ray and radio point source. 
                              In the upper right, a ClG is visible in X-rays, containing a radio source in the centre. 
                              Contours are: 0.3, 0.6, 1.2, 2.5, 5, and 10 mJy beam$^{-1}$, beam = 7\farcs05$\times$6\farcs63, $\lambda=20$ cm.
           {\it Upper right:} The upper radio source correlates with X-ray emission from source No 65.
                              The lower source is a ClG and not included in our X-ray catalogue, due to the extended emission \citep{2012A&A...545A.128H}.
                              Two radio jets are visible from one or possibly two AGN, presumably in the ClG.
                              Contours are: 0.3, 0.6, 1, 2, 3, 5, 7, 10, 15, and 20  mJy beam$^{-1}$, beam = 6\farcs56$\times$6\farcs16, $\lambda=20$ cm.
           {\it Lower left:}  Source No 27 is a galaxy candidate showing soft X-ray emission and an extended radio structure.
                              Contours are: 0.18, 0.3, 0.6, 1, 2, 3, 5 mJy beam$^{-1}$, beam = 6\farcs56$\times$6\farcs16, $\lambda=20$ cm.  
           {\it Lower right:} Source No 3 is a bright X-ray source in the centre of two radio lobes.
                              Contours are: 2 to 50 mJy beam$^{-1}$ in steps of 2 mJy beam$^{-1}$, beam = 17\farcs8$\times$12\farcs2, $\lambda=20$ cm.
          }
  \label{fig:images}
\end{figure*}

\section{Observations and data reduction}
\label{sec:data}

\subsection{The XMM-Newton survey of the SMC}
\label{sec:data:xray}

The observatory \xmm\ \citep{2001A&A...365L...1J} is equipped with three X-ray telescopes \citep{2002SPIE.4496....8A}, 
with EPIC CCD detectors \citep{2001A&A...365L..18S,2001A&A...365L..27T} in their focal planes. 
\xmm\ performed a survey of the SMC \citep{2012A&A...545A.128H}, completely covering the main body with a field size of 5.58 deg$^2$ and
a limiting sensitivity of $\sim$$2\times10^{-14}$ erg cm$^{-2}$ s$^{-1}$ in the (0.2--12.0) keV band.
For each \xmm\ observation, a maximum-likelihood source detection was performed on X-ray images of various energy bands simultaneously, i.e. a similar method as used for the 
\xmm\ serendipitous source catalogue \citep[][]{2009A&A...493..339W}. 
This resulted in a catalogue of 3053 X-ray sources 
and includes additional outer fields, leading to a total area of 6.32 deg$^2$. 
For the list of used observations and a 
detailed description of the X-ray point-source catalogue, see \citet{smcsrccat}.
Extended X-ray sources can be found in \citet{2012A&A...545A.128H}.

\subsection{Radio observations of the SMC}
\label{sec:data:radio}

Radio-continuum images used in this study (Table~\ref{tab:radio}) were created by combining data from the Australia Telescope Compact Array (ATCA) 
with data obtained from Parkes radio studies \citep{2002MNRAS.335.1085F,2011SerAJ.183...95C,2011SerAJ.182...43W,2011SerAJ.183..103W,2012SerAJ.184...93W}. 
We also used high-resolution images from \citet{2010SerAJ.181...63B} and newly created images of the N\,19 region \citep{2012SerAJ.185...53W}. 
To complement our study we included an image at 36~cm, which was obtained from the MOST survey \citep{1993LNP...416..167Y}.

\begin{table*}
 \centering
 \caption{Details of radio-continuum data and surveys used in this study.}
 \label{tab:radio}
  \begin{tabular}{lccccl}
   \hline\hline\noalign{\smallskip}
   \multicolumn{1}{l}{Column} &   
   \multicolumn{1}{c}{$\lambda$} &   
   \multicolumn{1}{c}{$\nu$} &   
   \multicolumn{1}{c}{Beam size} &   
   \multicolumn{1}{c}{R.M.S.} &  
   \multicolumn{1}{l}{Reference}\\
   Table~\ref{tab:sources} & (cm) & (MHz) & (arcsec) & (mJy/beam) & \\
   \noalign{\smallskip}\hline\noalign{\smallskip}
   (9) & 36 &  843 & 45$\times$45     & 1.5 & \citet{1993LNP...416..167Y}         \\
  (10) & 20 & 1420 & 17.8$\times$12.2 & 1.5 & \citet{2002MNRAS.335.1085F,2011SerAJ.182...43W,2011SerAJ.183..103W} \\
  (11) & 20 & 1420 & 7.0$\times$6.6   & 0.1 & \citet{1995MNRAS.275.1218Y,2001AJ....122..849D,2010SerAJ.181...63B,2012SerAJ.185...53W} \\
  (12) & 13 & 2370 & 40$\times$40     & 1.5 & \citet{2002MNRAS.335.1085F,2011SerAJ.183..103W}         \\
  (13) & 13 & 2370 & 7.0$\times$6.6   & 0.1 & \citet{1995MNRAS.275.1218Y,2001AJ....122..849D,2010SerAJ.181...63B,2012SerAJ.185...53W} \\
  (14) &  6 & 4800 & 30$\times$30     & 0.5 & \citet{2011SerAJ.183...95C,2012SerAJ.184...93W}     \\
  (15) &  3 & 8640 & 20$\times$20     & 0.5 & \citet{2011SerAJ.183...95C,2012SerAJ.184...93W}      \\
   \noalign{\smallskip}\hline\noalign{\smallskip}
  \end{tabular}
\end{table*}

\begin{figure*}
\sidecaption
  \includegraphics[width=12cm]{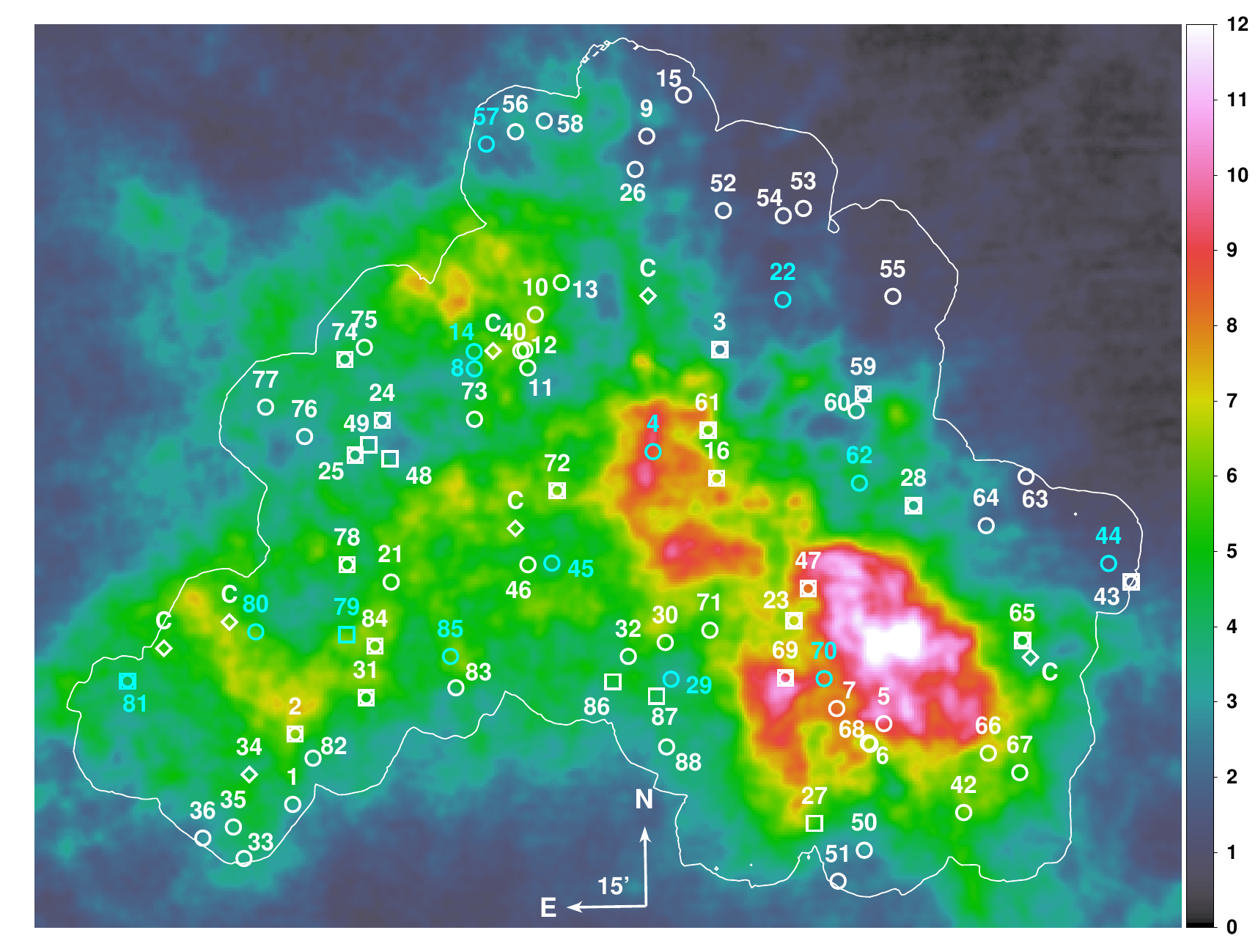}
  \caption{Spatial distribution of the X-ray sources with radio associations in the main field overplotted on the \HI\ map of \citet{1999MNRAS.302..417S}.
           The colour scale indicates the column density in units of $10^{21}$ cm$^{-2}$. 
           The white line marks the \xmm\ main field of the SMC survey. Seven additional sources are located in outer fields, not shown here. 
           Labels give the source number (Column 1 in Table~\ref{tab:sources}). Radio sources in X-ray clusters of galaxies (Table~\ref{tab:radio_clg}) are labelled with {\tt C}.
           X-ray radio correlations are shown by circles. If the source shows a jet-like structure in radio, it is shown by a boxed circle.
           Sources that are classified as galaxies are shown by boxes.
           Sources that are within ClGs are marked with diamonds. 
           Radio sources with $-0.3<\alpha<0$ are plotted in cyan, others in white.
          }
  \label{fig:spatial}
\end{figure*}

\subsection{Correlation}
\label{sec:data:correlation}
\addtocounter{table}{1}

Radio-continuum emission of background sources can show spatial structures caused by jets emitted from the AGN. 
Examples from this work are shown in Fig.~\ref{fig:images}.
In some cases, the X-ray source, which marks the position of the AGN candidate, is placed in between two radio jets.
To find all such sources, we visually inspected the radio images for counterparts of X-ray sources.
We found 88 out of the 3053 sources from the X-ray point-source catalogue with a counterpart visible in at least one radio image.
In Table~\ref{tab:sources}, we list the radio sources with an X-ray match. The columns give the following parameters:\\
(1) Running number, No, of the sources in this study;\\
(2) Source number from the \xmm\ SMC point-source catalogue \citep{smcsrccat};\\
(3) X-ray flux in the (0.2--4.5) keV band in $10^{-14}$ erg cm$^{-2}$ s$^{-1}$;\\
(4) Hardness ratio $HR_2=(R_3-R_2)/(R_3+R_2)$ with $R_2$ and $R_3$ being the X-ray count rates in the (0.5$-$1.0) and (1.0$-$2.0) keV band;\\
(5-6) Sexagesimal J2000 coordinates as derived from radio. 
For point-like sources a Gaussian fit was used to determine the position, whereas for complex jet-like structures, the position of the peak flux is given. 
In the case of two radio detections obviously comprising two jets of the same source we list the apparent centre of the perceived origin of the jets;\\
(7) Estimated position uncertainty for the radio position in arcsec based on image resolution;\\
(8-14) Integrated flux densities $S_\nu$ in mJy at various radio frequencies $\nu$ from the data described in Table~\ref{tab:radio};\\
(15) Radio spectral index $\alpha$ according to $S_\nu \sim \nu^\alpha$ and uncertainty following \citet[][]{2004MNRAS.355...44P};\\
(16) Source classification from this work;\\
(17) Comments on individual sources and references to other catalogues. Jet-like radio structures are noted.

\section{Results and discussion}
\label{sec:results}

\begin{figure}
  \resizebox{\hsize}{!}{\includegraphics[angle=0,clip=]{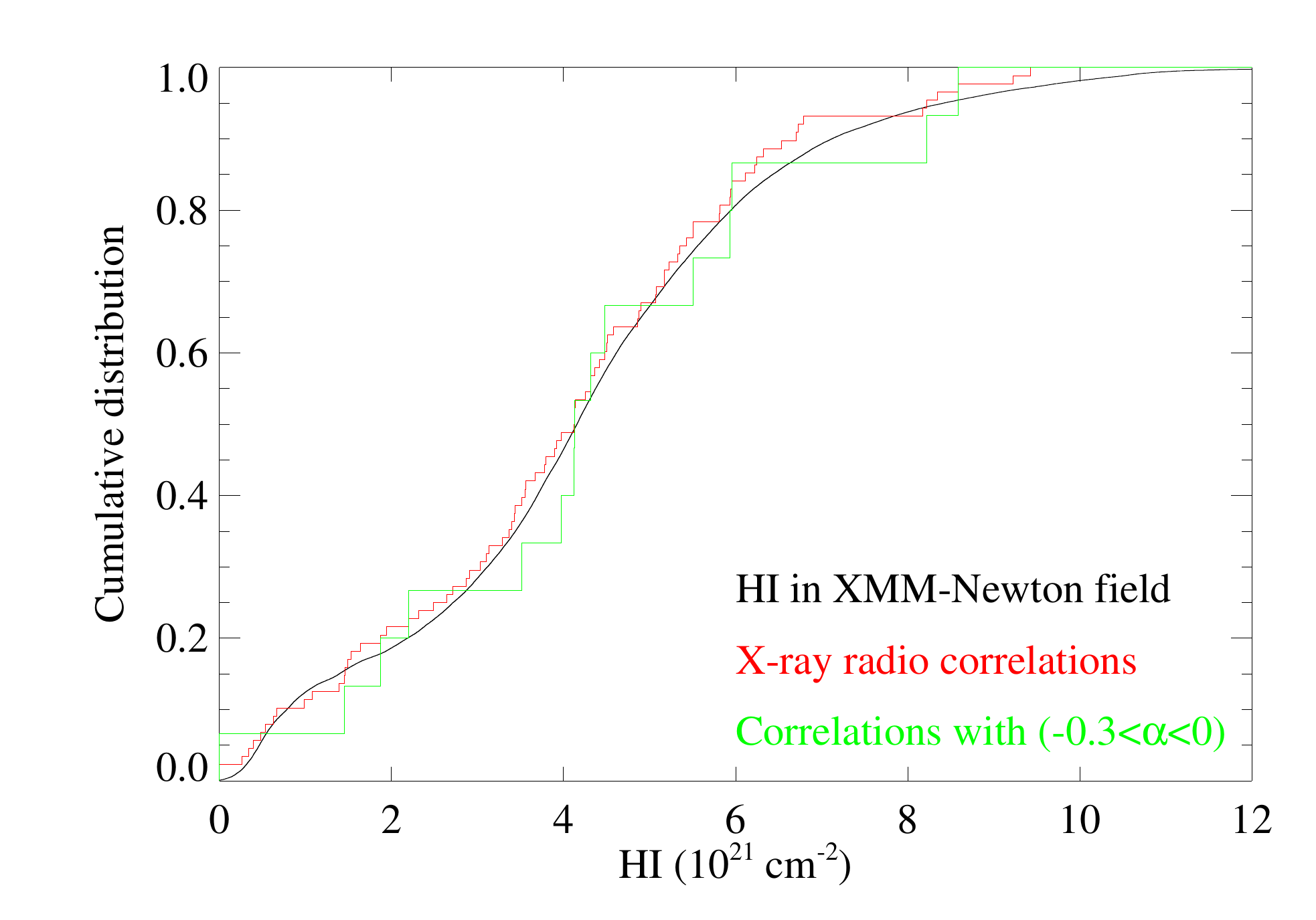}}
  \caption{Cumulative distribution of \ion{H}{i} column density in the \xmm\ field (black) and of the line-of-sight column densities of our X-ray radio correlations (red). Correlations with $-0.3<\alpha<0$ are plotted in green.}
  \label{fig:ks}
\end{figure}

The \xmm\ survey of the SMC provides a continuous coverage of the bar and eastern wing of the SMC and allowed the creation of the most comprehensive X-ray point-source catalogue.
At the same time, deeper radio images reveal fainter sources in an even larger area.
From the correlation of both datasets we derived a list of 88 discrete sources with X-ray and radio emission.
In Fig.~\ref{fig:spatial}, we mark all X-ray sources with radio counterparts within the main field (5.58 deg$^2$) of the \xmm\ survey on an SMC \ion{H}{i} image from \citet{1999MNRAS.302..417S}.
The \xmm\ main field is indicated with a white contour. For a description of additional fields disconnected from the main field see \citet{smcsrccat}.
We do not see a particular correlation of the source density with the SMC \ion{H}{i} intensity, which is consistent with the sources being mainly background objects.
This is also evident when comparing the relative line-of-sight \ion{H}{i} column density of our sources with the \ion{H}{i} distribution in the \xmm\ field (Fig.~\ref{fig:ks}).
As expected, both show a similar distribution, i.e. we do not find more sources in regions with higher (or lower) \ion{H}{i} as it would be the case for a correlation (or anti-correlation).
Forty-five of our sources are located behind dense SMC regions with an \ion{H}{i} column density of N$_{\rm H}^{\rm SMC}>4\times10^{21}$ cm$^{-2}$, six sources have a N$_{\rm H}^{\rm SMC}>8\times10^{21}$ cm$^{-2}$.
Also no structure according to the bar of the SMC is seen, where sources of star-forming regions are found
(e.g. compare compact \ion{H}{ii} regions of \citet[][their Fig. 10]{2012SerAJ.185...53W}
and YSO of \citet[][their Fig. 1]{2013MNRAS.428.3001O}.

\subsection{Correlation statistics}
\label{sec:results:chancecor}

\begin{figure}
  \resizebox{\hsize}{!}{\includegraphics[angle=-90,clip=]{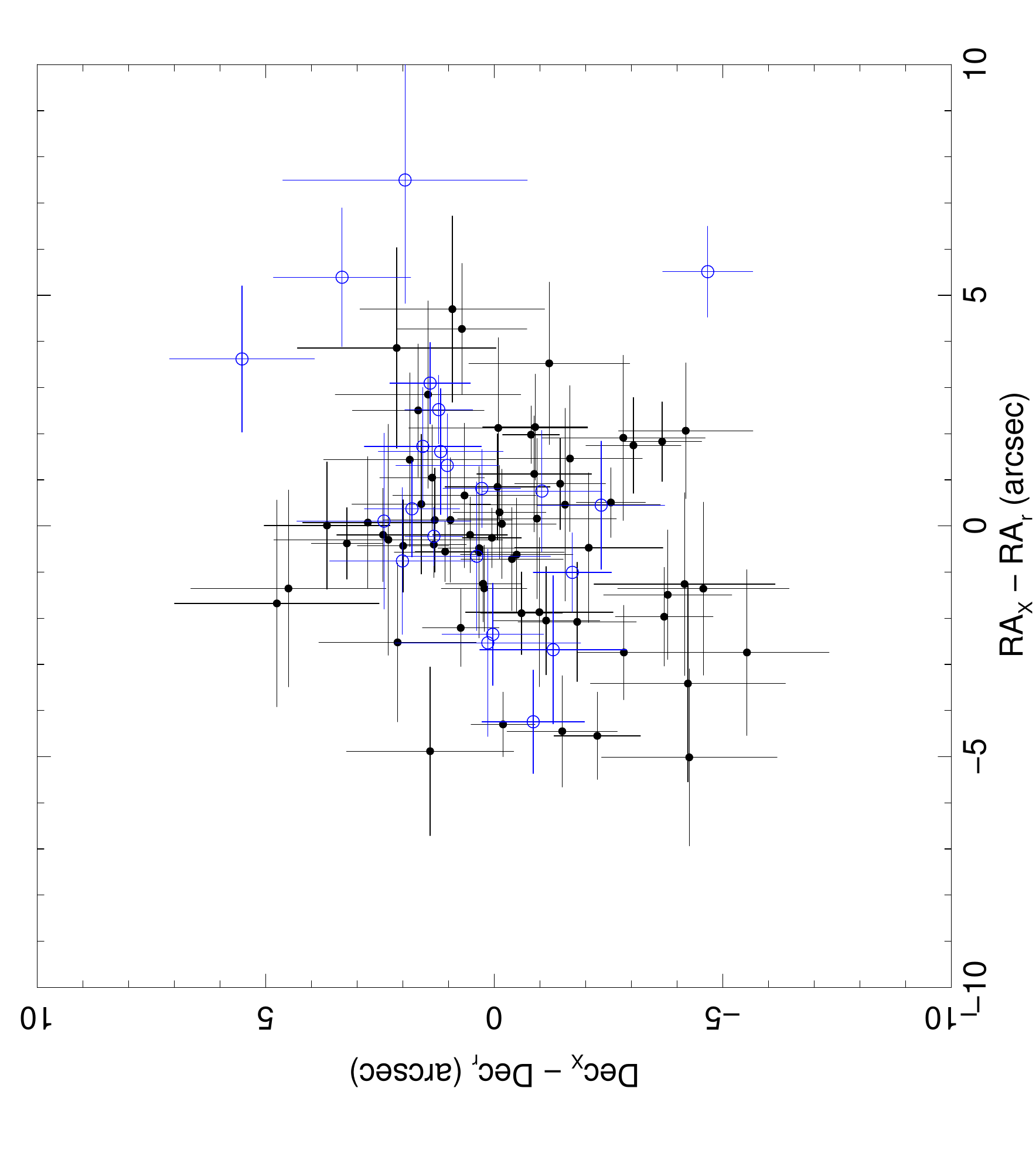}}
  \caption{Separation of X-ray and radio positions. Sources, which have some extent or jets in X-ray or radio are plotted with blue open circles, 
           other sources with black filled circles. Error bars mark 1$\sigma$ confidence.}
  \label{fig:sep}
\end{figure}

We show the angular separation between radio and X-ray positions in Fig.~\ref{fig:sep}. 
Eighty-four of our 88 correlations have an angular separation of $d\leq5\times (\sigma_{\rm r}^2 + \sigma_{\rm X}^2 )^{1/2}$, 
where $\sigma_{\rm r}$ and $\sigma_{\rm X}$ are the position uncertainties of the radio and X-ray source.
In some cases (e.g. source No 3, Fig.~\ref{fig:images} lower right), the small errors and the radio extent cause larger separations as in these cases, the position of the central radio source had to be estimated. 
The error-weighted average offset for point sources is $\Delta$RA=0\farcs14 and $\Delta$Dec=-0\farcs07 with an uncertainty of 0\farcs18. 
We do not see systematic deviations.

Since the source correlation was done manually, the determination of the fraction of chance coincidences is not straight forward.
However, we can compare our result with the result of a simple angular-separation-based cross matching.
We merged the radio catalogues of \citet[][$\lambda=13, 20, 36$ cm]{2011SerAJ.183..103W} and \citet[][$\lambda=3, 6$ cm]{2012SerAJ.184...93W}, as well as the 13 cm sources of \citet{2002MNRAS.335.1085F}.
These catalogues are based on the same radio data, but omit the deep images and therefore contain only 60 of our 89 radio sources.
Also, some sources with long jets were rejected for the radio point-source catalogues.
We find 58 of our X-ray radio associations within $d\leq3.439\times (\sigma_{\rm r}^2 + \sigma_{\rm X}^2 )^{1/2}$.
Additional 6 correlations were not selected in our manual correlation, as these were regarded as too uncertain, e.g. if a jet-like structure is not pointing towards the X-ray source.
To check for chance coincidences, we shifted the coordinates of one catalogue by $\gtrsim$50\arcsec\ in different directions.
This resulted in 2.5$\pm$1.6 correlations indicating a chance coincidence rate of $\sim$5\%.

\subsection{Spectral characteristics}
\label{sec:results:spectral}

\begin{figure}
  \resizebox{\hsize}{!}{\includegraphics[angle=0,clip=]{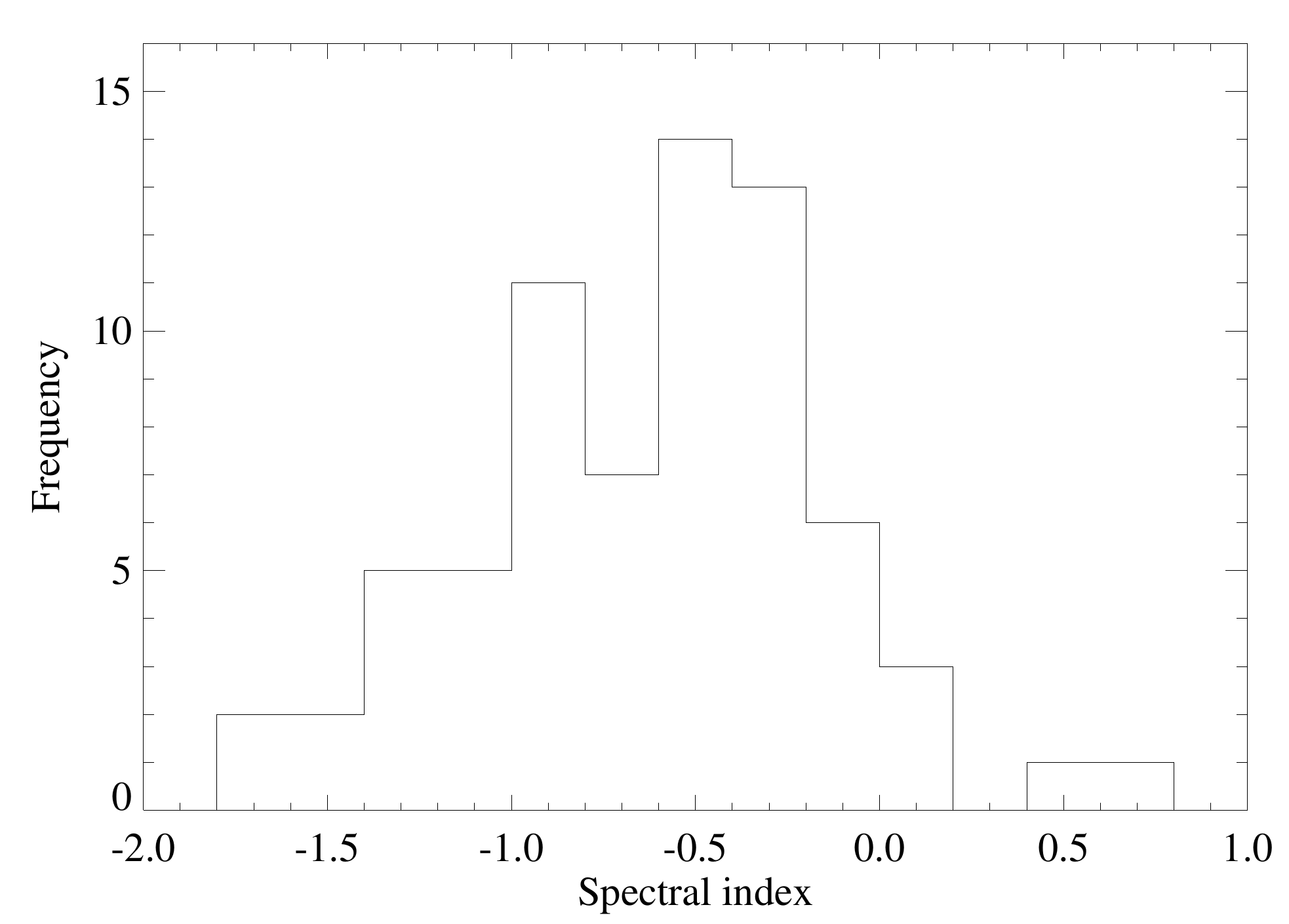}}
  \caption{Histogram of all sources with determined radio spectral index $\alpha$. The bin width is 0.2.
          }
  \label{fig:specind}
\end{figure}

The spectral index $\alpha$ of radio background sources covers a wide range, but is on average steeper for background sources than for SNRs or the thermal radio emission from \HII\ regions \citep[c.f.][]{1998A&AS..130..421F}. 
For other rare source types, see the following sections.
Because the possible $\alpha$ values of AGN, SNRs, and \HII\ regions are overlapping, the classification of radio sources based only on $\alpha$ is ambiguous, 
but with the detection of X-rays strongly points to a background object. 
The distribution of the spectral index $\alpha$ as estimated from the radio images is shown in Fig.~\ref{fig:specind}. 
As expected we find a relatively wide distribution in our sample. 
60\% of the 70 sources with determined spectral index have a steep spectrum ($\alpha<-0.45$ with $S_\nu \sim \nu^\alpha$) and 16\% show a flat spectrum ($\alpha>-0.2$). 
Compared to an unbiased radio sample of background sources \citep[][]{2004MNRAS.355...44P}, we find more sources with flat spectrum due to the X-ray selection of our sample \citep{1994A&AS..106..303N}.

Twenty-five sources (36\%) have a very steep spectral index of $\alpha<-0.8$ and are excellent candidates for compact steep spectrum (CSS) sources \citep{1998PASP..110..493O,2009AN....330..120F}.
CSS sources are believed to bridge the evolutionary phase between the early gigahertz peaked-spectrum (GPS) sources and later and larger Fanaro-Riley Type~I and II (FR~I/II) galaxies. 
Sources No 57 and 70 are perfect candidates for GPS sources, due to  their curved spectral index.

Radio sources with a rather flat radio spectrum and no indication of a radio jet or extended structure in X-rays are good candidates for BL\,Lac objects.
The nearly featureless spectrum of these sources makes them ideal background emitters to measure absorption effects of the interstellar medium in the SMC.
We find 22 compact radio sources that have an $\alpha>-0.5$ and are not classified as foreground object.
These sources are good candidates for BL\,Lac objects.

The inverse spectral index ($\alpha>0$) of source No 28 can be explained by radio variability due to non-simultaneous measurements, 
as seen from the different flux densities measured at 20~cm. This can also be the case for sources No 26, 30, and 73.

\begin{figure}
  \resizebox{\hsize}{!}{\includegraphics[angle=0,clip=]{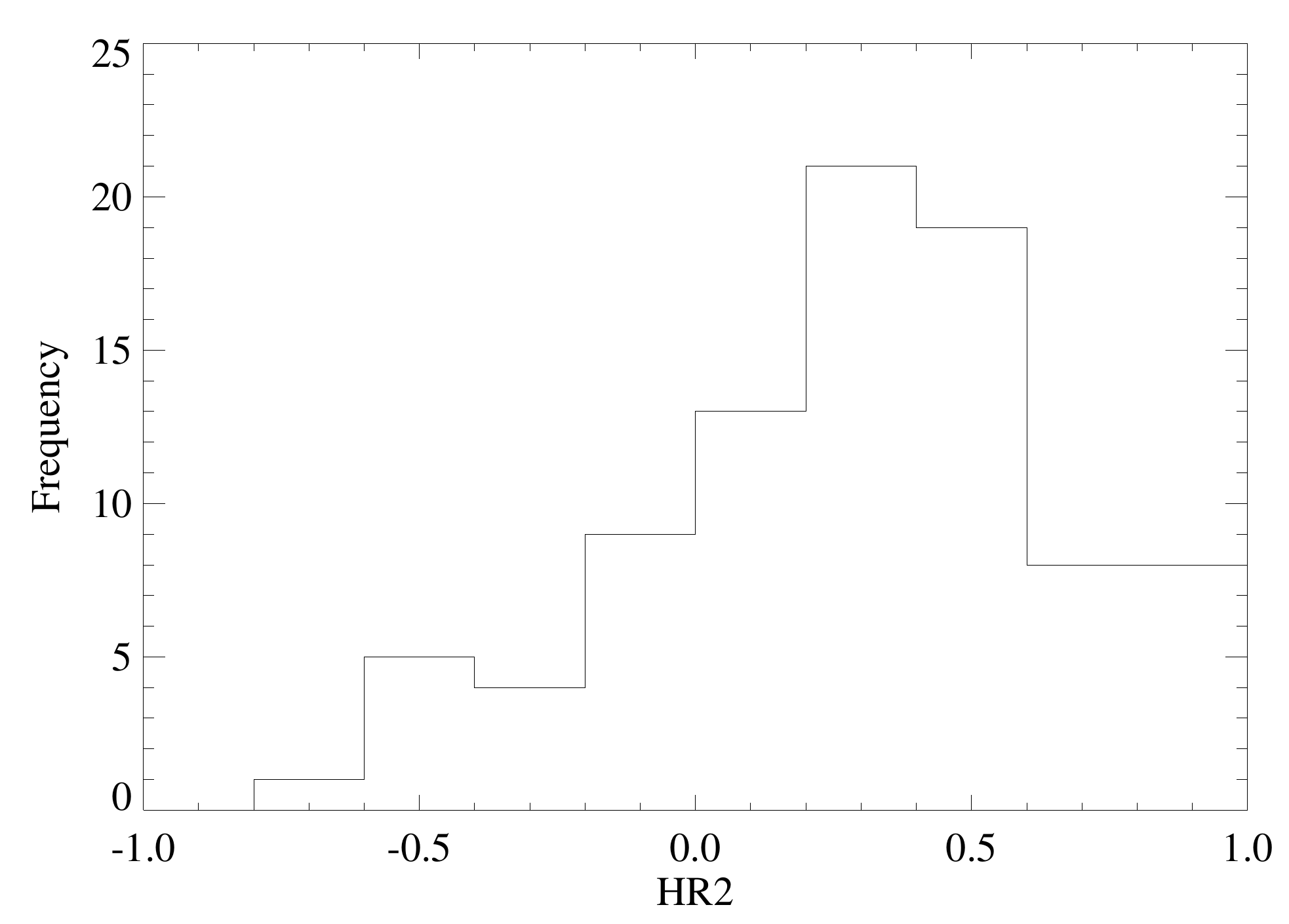}}
  \caption{Histogram of the X-ray hardness ratio $HR_2$ distribution. The bin width is 0.2.
          }
  \label{fig:hr2}
\end{figure}

In Fig.~\ref{fig:hr2}, we show the distribution of $HR_2$, defined by $(R_3-R_2)/(R_3+R_2)$, where $R_2$ and $R_3$ is the X-ray count rate in the (0.5$-$1.0)~keV and (1.0$-$2.0)~keV band, respectively.
Hard X-ray sources, such as AGN or X-ray binaries, show higher $HR_2$ values than soft X-ray sources such as Galactic stars or SNRs.
For AGN, typical values are $HR_2>0$ 
, whereas most foreground stars are found with $HR_2<0$ \citep[e.g. ][]{2011A&A...534A..55S}.
Normal galaxies can show soft and hard X-ray emission, depending on the contribution from X-ray binaries, hot interstellar medium and SNRs.
As expected for background sources, the X-ray to radio correlations mainly show hard X-ray emission. 

\subsection{Galaxies}
\label{sec:results:galaxies}

Six of our sources (No 17, 18, 24, 48, 79, and 87) have a counterpart in the 6dF galaxy survey with determined redshifts \citep{2009MNRAS.399..683J}. 
Remarkably, all of the 6dF correlations with the 3053 X-ray sources show radio emission. No 17 is a known Seyfert 2 galaxy.
All these sources have an entry in the 2MASS extended source catalogue \citep[2MASX,][]{2006AJ....131.1163S}. 
In addition we find 2MASX counterparts for sources No 27 and 86, pointing to a galaxy nature of these sources.
The galaxy nature is further supported by a 2MASX galaxy score of 1.0 and a $HR_2<0$.
Source No 27 is shown in Fig.~\ref{fig:images}, lower left.

\subsection{Galactic stars}
\label{sec:results:stars}

Galactic stars can show X-ray and radio emission.
The correlation No 37 is identified with the Galactic star CF\,Tuc.
This is an active RS CVn-type binary with known X-ray and radio emission \citep{1996A&A...311..211K,1997MNRAS.287..199G}.
In addition to soft X-ray emission from coronal plasma, a bright ($\log{(F_{\rm X}/F_{\rm opt})} < -1 $) optical counterpart is expected for a foreground star \citep{1988ApJ...326..680M}.

Eight of our sources show soft X-ray emission, compatible with the hardness-ratio criteria for stars of \citet{2004A&A...426...11P} or \citet{smcsrccat}
and have an appropriate optical counterpart in the Magellanic Cloud Photometric Survey \citep[MCPS, ][]{2002AJ....123..855Z}.
Five of these can be rejected as stars, because of a 2MASX counterpart (see Sec.~\ref{sec:results:galaxies}).
No 14 is along the SNR DEM\,S128, which causes a softer $HR_2$. However, in the X-ray image, a hard X-ray source is clearly seen.
Because we are looking out of the Galactic plane we expect a low contribution of young Galactic stars and 
the jet-like structure of No 47 is more likely caused by a background AGN than by outflows of protostellar jets of a Galactic source.
Source No 54 cannot be excluded as possible foreground star.
All other sources show hard X-ray emission that is not expected from coronal stellar X-ray emission.

We do not expect a contribution of SMC stars in the X-ray sample, 
because the moderate X-ray emission of normal stars even during flares \citep[$L_{\rm X}\lesssim 10^{33}$ erg s$^{-1}$, ][]{2009A&ARv..17..309G,2002ASPC..277..115F} is below the detection limit of the \xmm\ survey.
Further, only the brightest YSOs are detected in the radio observations \citep{2013MNRAS.428.3001O}.

\subsection{Cluster of galaxies}
Often, clusters of galaxies (ClGs), seen in X-rays, contain radio-continuum sources \citep[e.g.][and references therein]{2009A&A...501..835M}.
In these cases, the X-ray emission is caused by the hot intracluster medium, whereas the radio-continuum emission originates in an AGN in or in the direction of the ClG.
The sources No 34 and 72 were fitted with a significant extent in X-rays (11\farcs9$\pm$1\farcs1 and 12\farcs6$\pm$1\farcs6) and show hard X-ray emission, which points to a ClG nature of these sources.
We found also strong radio jets in other cluster candidates, which have a larger extent and were therefore not included in the X-ray point-source catalogue. 
These cluster candidates can be found in \citet{2012A&A...545A.128H}. 
The radio counterparts are listed in Table~\ref{tab:radio_clg}.
Examples are given in both upper images of Fig.~\ref{fig:images}.

\begin{table*}
  \centering
 \caption{Radio sources in X-ray selected clusters of galaxies from \citet{2012A&A...545A.128H}.}
 \label{tab:radio_clg}
  \begin{tabular}{rrrlllllllcr}
   \hline\hline\noalign{\smallskip}
     \multicolumn{1}{c}{RA$_{\rm r}$} &
     \multicolumn{1}{c}{Dec$_{\rm r}$} &
     \multicolumn{1}{c}{ePos} &
     \multicolumn{1}{c}{$S_{0.843}$} &
     \multicolumn{1}{c}{$S_{1.42}$} &
     \multicolumn{1}{c}{$S_{2.37}$} &
     \multicolumn{1}{c}{$S_{4.8}$} &
     \multicolumn{1}{c}{$S_{8.64}$} &
     \multicolumn{1}{c}{$\alpha$}  &
     \multicolumn{1}{c}{comment}  \\
   \noalign{\smallskip}\hline\noalign{\smallskip}
00 42 03.3 & -73 07 23  & 1.2  & 197.4  & 139.23    & 92.0 & 37.6 & 25.2 & -0.92 $\pm$0.07 &  H00, P04, jet  \\
00 58 22.6 & -72 00 45  & 0.6  & 21.2   & 18.8      & 9.9  & 5.2  & --   & -0.83 $\pm$0.07 &  H00, P04, jet  \\ 
01 04 07.0 & -72 43 51  & 1.1  & 19.3   & 11.3      & 8.6  & 3.6  & 2.2  & -0.93 $\pm$0.07 &  H00, P04       \\
01 04 45.6 & -72 10 19  & 0.5  & --     & 0.65 (HR) & --   & --   & --   & --              &  H00            \\    
01 16 31.3 & -72 58 02  & 1.5  & 2.1    & 2.2       & 1.9  & --   & --   & -0.10 $\pm$0.10 &                 \\
01 19 26.0 & -73 01 46  & 1.9  & 4.5    & 2.9       & 1.9  & --   & --   & -0.79 $\pm$0.05 &  jet            \\
   \noalign{\smallskip}\hline\noalign{\smallskip}
  \end{tabular}
\tablefoot{Values and references analogous to Table~\ref{tab:sources}.}
\end{table*}

\subsection{X-ray binaries}
Because of the relatively small stellar mass of the SMC, only a few low-mass X-ray binaries (LMXBs) are expected and none is known to date \citep{2010ASPC..422..224C}.
In the case of LMXBs, we would not expect to find an optical counterpart.
Also no ultra-luminous X-ray source (ULX) in the SMC is known 
and the measured X-ray luminosities are well below the ULX definition ($>$$10^{39}$ erg s$^{-1}$).
Therefore, the presence of a microquasar in our sample is unlikely, but cannot be ruled out.
Redshift measurements of AGN candidates are needed, to further reduce the possibility of a microquasar in the SMC, 
that otherwise might only be recognised during a bright outburst.

The SMC hosts around one hundred known high-mass X-ray binaries (HMXB).
Only one source in our sample has a proper optical counterpart in the MCPS that is compatible with an early-type star of a HMXB.
This source, No 5 = \object{[SG2005]\,SMC\,34},
was classified as ``HMXB candidate'' from \xmm\ and optical data by \citet[][ their source 34]{2005MNRAS.362..879S} and as ``new HMXB'' from Chandra observations by \citet[][source 4\_4]{2009ApJ...697.1695A}. \citet[][]{2009ApJ...697.1695A} also found additional possible optical counterparts from OGLE \citep{1998AcA....48..147U} in agreement with the X-ray position. \citet[][see their Fig. 2]{2011ApJS..194...22K} spectroscopically identified one of the fainter counterparts as a quasar with emission lines at redshift of $z=0.108$. The Chandra source of \citet[][]{2009ApJ...697.1695A} has a similar angular separation to both OGLE sources, the star (SMC-SC\,159896, 0.54\arcsec) and the quasar (SMC-SC\,159964, 0.53\arcsec). Our radio source has an angular separation of 0.60\arcsec and 0.58\arcsec, respectively, with a position uncertainty of 1\arcsec.
From a spectral analysis of the \xmm\ spectrum we could derive rough parameters for an absorbed power-law model. The photon index is between $\Gamma = (0.99-1.97)$ and the source is highly absorbed with \nh = $(1.7-3.9)\times 10^{23}$ cm$^{-2}$. This is compatible with both AGN and HMXB, which have typical photon indices of 1.75 \citep{2006A&A...451..457T} and 1.0 \citep{2004A&A...414..667H}, respectively, and can show high intrinsic absorption. Therefore, the radio emission is likely caused by the quasar and the X-ray emission can be caused by both, however a background object is more likely.

\subsection{Pulsar wind nebulae}
A rare but interesting source class are PWNe \citep[e.g.][]{2006ARA&A..44...17G}.
PWN candidates in the SMC are X-ray and radio emitting sources that are found within thermal SNR shells, like the central sources of HFPK\,334 \citep{2008A&A...485...63F} and of IKT\,16 \citep{2011A&A...530A.132O}.
In the Large Magellanic Cloud (LMC), five candidates for such composite SNRs are known \citep[B0540-693, N157B, B0532-710, DEM\,L241, and SNR\,J0453–6829, see][and references therein]{2012A&A...543A.154H}.
However, also PWNe without a thermal SNR shell, like the Crab Nebula or the PWN around the pulsar PSR\,B0540-69 in the LMC, are possible.
The fact, that we do not know such a system in the SMC might be a selection effect.

The expected X-ray spectra of PWNe ($\Gamma\sim2$) will result in similar hardness ratios as for AGN ($\Gamma\sim1.75$) and hamper an X-ray-based classification.
Typical spectral radio indices of PWNe are $-0.3<\alpha<0$, where we find 15 sources in our sample. 
These sources might be taken as a flux limited sample of candidates for PWNe.
However, the source distribution (plotted in cyan in Fig.~\ref{fig:spatial} and green in Fig.~\ref{fig:ks}) is not correlating with star-forming regions in the SMC, i.e. the bar of the SMC where we see most of the SNRs.
Therefore, we expect that most of these sources are background objects and that there is no significant contribution of PWNe given the sensitivity of our observations.

A special case is source No 14, which is found along the SNR \object{DEM\,S128} (upper left of Fig.~\ref{fig:images}) with a spectral index of $\alpha=-0.21\pm0.12$, 
which is steeper than for the surrounding SNR \citep[$\alpha=-0.48\pm0.06$, ][]{2000A&A...353..129F}.
The type-Ia classification of the SNR \citep{2004A&A...421.1031V} and the offset of its X-ray bright centre from the point source suggest that both sources are not connected with each other.
In the SAGE survey \citep{2011AJ....142..102G} we find a mid-infrared Spitzer/IRAC counterpart (SSTISAGEMA\,J010530.69-721021.3) with colours that might be consistent with an AGN. 
However, in the deep X-ray images, the emission from DEM\,S128 clearly extends further towards the north-west \citep[compare also ][Fig~6.1]{2012A&A...545A.128H}. 
Therefore, we cannot exclude No 14 as a candidate for a PWN. DEM\,S128 will be discussed in more detail by Roper et al. (in prep.).

We can also roughly estimate the probability of finding a radio X-ray association within an SNR using our source list that contains 2.4 sources deg$^{-2}$ with $-0.3<\alpha<0$.
According to \citet{2012A&A...545A.128H}, the area covered by known SNRs is $\sim$0.044 deg$^{-2}$.
For a Poisson distribution, we would expect to find one source along an SNR with a likelihood of 9.4\% and more than one source with a likelihood of 0.5\%.
So the DEM\,S128 correlation might still be by chance, but this is unlikely to be the case for all PWN candidates in the SMC (HFPK\,334 and IKT\,16, not included in this study).

\subsection{Comparison with previous studies}

For 17 of our sources, we find a counterpart in the ROSAT PSPC catalogue \citep[][]{2000A&AS..142...41H}.
These are commented with H00 in Table~\ref{tab:sources}. 
Twelve of these sources were already associated with radio sources at $\lambda=13$~cm in that work.
In addition, the ROSAT catalogue lists 39 sources with radio association.
Of these, nine are outside the \xmm\ field ([HFP2000] 52, 124, 138, 347, 357, 522, 685, 687, and 692),
four are now resolved as ClGs ([HFP2000] 101, 147, 317, and 410) 
and 18 are SNRs ([HFP2000] 45, 107, 125, 145, 148, 194, 217, 281, 285, 334, 401, 413, 414, 419, 437, 454, 461, and 530), where
[HFP2000] 281 was probably detected as part of a  bubble connected to the SNR DEM\,S68 \citep[][]{2008A&A...485...63F}.
[HFP2000] 88 was not detected with \xmm.
For [HFP2000] 49, 206, 249, 380, 440, 448, and 668, the improved X-ray positions with respect to ROSAT make an X-ray radio association unlikely.

Since the gain of the radio data is rather in sensitivity than in resolution, the comparison with \citet[][]{2004MNRAS.355...44P} is somewhat more straight forward.
We find 32 of our sources in \citet[][]{2004MNRAS.355...44P}, marked with P04 in Table ~\ref{tab:sources}.
All sources but one were classified as background objects by these authors, but only 3 sources are noted as X-ray association.
Only source No 22 (\object{[FBR2002] J005254-720132}) was classified as background object or \HII\ region. The X-ray emission points to a background source in this case.
Further, there are 37 sources for which \citet{2004MNRAS.355...44P} give an X-ray luminosity or an X-ray comment.
Of these, 13 were identified as SNRs,
three are in ClGs, and
14 were classified as \HII\ regions where we do not find X-ray sources in our study, confirming the classification.
From the remaining sources one is classified as ``XRB'' (microquasar candidate), correlating with [HFP2000] 295. This X-ray radio association was rejected above.
The other sources are candidate background objects, but we do not find X-ray counterparts in the \xmm\ catalogue
([FBR2002]
J004552-731339,
J004836-733056,
J005218-722708,
J005602-720908,
J005610-721833, and
J010525-722525).

\section{Conclusions}
\label{sec:conclusions}

We inspect the positions of X-ray sources from the \xmm\ SMC survey in the corresponding deep radio-continuum images and found 88 X-ray sources associated with a unique radio counterpart.
One source is identified with a foreground star, 
one is a confirmed quasar probably confused with a HMXB candidate, 
eight are identified or classified as galaxies, 
two radio sources are within clusters of galaxies,
and one might be a PWN.
The remaining 75 X-ray sources associated with a unique radio counterpart are classified as AGN behind the SMC.
Due to the precise X-ray positions of our X-ray catalogue ($\sim$1.5\arcsec) and the low density of radio sources in the SMC field, chance correlations are unlikely and we derive a high purity for our sample.
We expect the contribution of stars to our sample to be $\leq2$.
From background source candidates, seven are infra-red-selected candidates of \citet{2009ApJ...701..508K}, 
31 were classified as background radio sources by \citet{2004MNRAS.355...44P}, 
and 11 as AGN candidates by \citet{2000A&AS..142...41H}. 
40 associations are newly classified background objects behind the SMC, 
for the others, the X-ray radio association affirms the previous background-object classification. 
For a total of 21 X-ray point sources, we find a jet like structure in radio, which points to the AGN character of the source. 
In addition, we list six radio sources inside ClGs with high X-ray extent, where three radio sources show a jet.

\begin{acknowledgements}
The XMM-Newton project is supported by the Bundesministerium f\"ur Wirtschaft und 
Technologie/Deutsches Zentrum f\"ur Luft- und Raumfahrt (BMWI/DLR, FKZ 50 OX 0001)
and the Max-Planck Society. 
The ATCA is part of the Australia Telescope which is funded by the Commonwealth of Australia for operation as a National Facility managed by CSIRO.
We used the {\sc karma} and {\sc miriad} software package developed by the ATNF. 
R.S. acknowledges support from the BMWI/DLR grant FKZ 50 OR 0907. 
\end{acknowledgements}

\bibliographystyle{aa}
\bibliography{../auto,../general}

%
%
\longtabL{2}{ \tiny
\begin{landscape}
\begin{longtable}{rcccccrrrrrrrcccl}
 \caption{\label{tab:sources} X-ray to radio-continuum correlations in the SMC field.}\\
     \hline\hline\noalign{\smallskip}
     \multicolumn{1}{c}{(1)} &
     \multicolumn{1}{c}{(2)} &
     \multicolumn{1}{c}{(3)} &
     \multicolumn{1}{c}{(4)} &
     \multicolumn{1}{c}{(5)} &
     \multicolumn{1}{c}{(6)} &
     \multicolumn{1}{c}{(7)} &
     \multicolumn{1}{c}{(8)} &
     \multicolumn{1}{c}{(9)} &
     \multicolumn{1}{c}{(10)} &
     \multicolumn{1}{c}{(11)} &
     \multicolumn{1}{c}{(12)} &
     \multicolumn{1}{c}{(13)} &
     \multicolumn{1}{c}{(14)} &
     \multicolumn{1}{c}{(15)} &
     \multicolumn{1}{c}{(16)} &
     \multicolumn{1}{c}{(17)} \\
     \multicolumn{1}{c}{No} &
     \multicolumn{1}{c}{ID$_{\rm X}$} &
     \multicolumn{1}{c}{$F_{\rm X}$} &
     \multicolumn{1}{c}{$HR_2$} &
     \multicolumn{1}{c}{RA$_{\rm r}$} &
     \multicolumn{1}{c}{Dec$_{\rm r}$} &
     \multicolumn{1}{c}{ePos} &
     \multicolumn{1}{c}{$S_{0.843}$} &
     \multicolumn{1}{c}{$S_{1.42}$} &
     \multicolumn{1}{c}{$S_{1.42}$} &
     \multicolumn{1}{c}{$S_{2.37}$} &
     \multicolumn{1}{c}{$S_{2.37}$} &
     \multicolumn{1}{c}{$S_{4.8}$} &
     \multicolumn{1}{c}{$S_{8.64}$} &
     \multicolumn{1}{c}{$\alpha$} &
     \multicolumn{1}{c}{class} &
     \multicolumn{1}{c}{comment\tablefootmark{a}} \\
\hline
\endfirsthead
\caption{Continued.} \\
     \multicolumn{1}{c}{No} &
     \multicolumn{1}{c}{ID$_{\rm X}$} &
     \multicolumn{1}{c}{$F_{\rm X}$} &
     \multicolumn{1}{c}{$HR_2$} &
     \multicolumn{1}{c}{RA$_{\rm r}$} &
     \multicolumn{1}{c}{Dec$_{\rm r}$} &
     \multicolumn{1}{c}{ePos} &
     \multicolumn{1}{c}{$S_{0.843 }$} &
     \multicolumn{1}{c}{$S_{1.42 }$} &
     \multicolumn{1}{c}{$S_{1.42}$} &
     \multicolumn{1}{c}{$S_{2.37}$} &
     \multicolumn{1}{c}{$S_{2.37}$} &
     \multicolumn{1}{c}{$S_{4.8 }$} &
     \multicolumn{1}{c}{$S_{8.64}$} &
     \multicolumn{1}{c}{$\alpha$} &
     \multicolumn{1}{c}{class} &
     \multicolumn{1}{c}{comment\tablefootmark{a}} \\
\hline
\endhead
\hline
\endfoot
 1 & 26    &  1.1$\pm$0.2   &   0.24 $\pm$ 0.13  & 01 14 27.85 & -73 33 12.7 & 1.5 & 8.68    & 1.97   & --     & 2.60   & --    & 3.21   & --     & -0.42$\pm$0.15  & AGN        &                                 \\ 
 2 & 32    &  7.7$\pm$0.3   &   0.56 $\pm$ 0.03  & 01 14 05.16 & -73 20 06.9 & 1.5 & 128.02  & 78.44  & --     & 51.93  & --    & 30.64  & 17.34  & -0.84$\pm$0.05  & AGN        & jet, P04, H00, K09              \\ 
 3 & 59    &  25.9$\pm$0.5  &   0.19 $\pm$ 0.02  & 00 55 23.16 & -72 10 55.8 & 1.5 & 242.21  & 76.68  & --     & 67.44  & --    & 46.78  & 28.90  & -0.79$\pm$0.05  & AGN        & jet, H00                        \\ 
 4 & 85    &  2.0$\pm$0.2   &   0.76 $\pm$ 0.16  & 00 58 15.42 & -72 30 07.5 & 1.5 & --      & 2.30   & --     & --     & --    & 2.25   & --     & -0.02$\pm$0.25  & AGN        & H00, K09                        \\ 
 5 & 117   &  5.4$\pm$0.3   &   0.53 $\pm$ 0.17  & 00 48 18.79 & -73 21 00.0 & 1.0 & --      & --     & 0.87   & --     & 0.45  & --     & --     & -1.33$\pm$0.25  & AGN/HMXB?  & K11, S05, K09                   \\ 
\noalign{\smallskip}                                                                                                                                                 
 6 & 118   &  5.3$\pm$0.3   &   0.31 $\pm$ 0.40  & 00 48 54.42 & -73 24 57.2 & 1.0 & 1.45    & --     & 1.65   & --     & 0.97  & --     & --     & -0.40$\pm$0.20  & AGN        &                                 \\ 
 7 & 135   &  0.6$\pm$0.1   &   0.23 $\pm$ 0.49  & 00 50 23.69 & -73 18 20.8 & 1.0 & --      & --     & 0.23   & --     & --    & --     & --     & --              & AGN        &                                 \\   
 8 & 186   &  4.0$\pm$0.1   &   0.20 $\pm$ 0.03  & 01 05 32.93 & -72 13 30.6 & 1.0 & 10.37   & 11.90  & 11.76  & 5.56   & 13.11 & 9.47   & 6.52   & -0.21$\pm$0.05  & AGN        & H00, P04                        \\ 
 9 & 217   &  3.3$\pm$0.3   &   0.49 $\pm$ 0.20  & 00 58 20.53 & -71 30 40.9 & 1.5 & 38.46   & 32.30  & --     & 12.27  & --    & 12.65  & 3.87   & -0.93$\pm$0.07  & AGN        & P04                             \\ 
10 & 236   &  2.7$\pm$0.1   &   0.81 $\pm$ 0.16  & 01 02 58.90 & -72 03 47.8 & 1.0 & --      & --     & 5.60   & --     & 4.00  & --     & --     & -0.68$\pm$0.25  & AGN        & K09                             \\ 
\noalign{\smallskip}                                                                                                                                                 
11 & 299   &  1.0$\pm$0.1   &   0.61 $\pm$ 0.28  & 01 03 21.24 & -72 13 44.5 & 1.0 & --      & 3.87   & 9.12   & 1.75   & 5.66  & 1.93   & --     & -0.96$\pm$0.17  & AGN        &                                 \\ 
12 & 302   &  0.4$\pm$0.1   &   0.37 $\pm$ 0.16  & 01 03 27.18 & -72 10 22.0 & 1.0 & --      & --     & 0.42   & --     & --    & --     & --     & --              & AGN        &                                 \\  
13 & 306   &  0.3$\pm$0.1   &   0.96 $\pm$ 0.22  & 01 01 53.07 & -71 57 57.8 & 1.0 & 2.45    & 1.27   & 2.58   & --     & 2.47  & 0.98   & --     & -0.44$\pm$0.18  & AGN        &                                 \\ 
14 & 316   &  0.5$\pm$0.2   &  -0.58 $\pm$ 0.20  & 01 05 30.91 & -72 10 21.0 & 1.0 & --      & 2.26   & 4.53   & 4.02   & 4.01  & 3.33   & 2.04   & -0.21$\pm$0.12  & AGN/PWN?   & along SNR DEM\,S128             \\ 
15 & 329   &  2.0$\pm$0.2   &   0.13 $\pm$ 0.10  & 00 56 52.60 & -71 23 00.3 & 1.5 & 120.56  & 79.17  & --     & 48.71  & --    & 20.94  & 8.92   & -1.12$\pm$0.05  & AGN        & P04                             \\ 
\noalign{\smallskip}                                                                                                                                                 
16 & 341   &  1.8$\pm$0.2   &   0.67 $\pm$ 0.09  & 00 55 36.15 & -72 35 14.1 & 1.5 & 90.00   & 19.94  & --     & 21.53  & --    & 12.92  & 10.54  & -0.87$\pm$0.05  & AGN        & jet, H00                        \\ 
17 & 365   &  72.9$\pm$1.2  &   0.32 $\pm$ 0.04  & 00 53 56.23 & -70 38 04.8 & 1.5 & 48.70   & 40.72  & --     & 11.35  & --    & --     & 3.45   & -1.20$\pm$0.07  & AGN/galaxy & Seyfert 2 galaxy, J09, S06      \\ 
18 & 376   &  2.2$\pm$0.2   &  -0.59 $\pm$ 0.08  & 00 52 34.56 & -70 28 16.8 & 1.5 & 31.90   & 31.86  & --     & --     & --    & --     & --     & 0.00$\pm$0.20   & galaxy     & J09, S06                        \\ 
19 & 381   &  1.5$\pm$0.3   &   0.09 $\pm$ 0.15  & 00 56 08.11 & -70 38 47.0 & 1.5 & 150.39  & 93.74  & --     & 30.56  & --    & --     & 7.61   & -1.32$\pm$0.08  & AGN        & P04                             \\ 
20 & 400   &  0.8$\pm$0.2   &  -0.56 $\pm$ 0.24  & 00 53 47.98 & -70 46 01.3 & 1.5 & --      & 4.26   & --     & --     & --    & --     & --     & --              & AGN        &                                 \\  
\noalign{\smallskip}                                                                                                                                                 
21 & 402   &  8.4$\pm$0.3   &   0.89 $\pm$ 0.04  & 01 09 29.90 & -72 52 46.1 & 1.5 & 4.64    & 1.98   & --     & --     & --    & --     & --     & -1.68$\pm$0.03  & AGN        & K09                             \\ 
22 & 449   &  2.0$\pm$0.2   &   0.22 $\pm$ 0.08  & 00 52 54.23 & -72 01 32.4 & 1.5 & 13.44   & 10.14  & --     & 9.73   & --    & 7.10   & 7.73   & -0.25$\pm$0.13  & AGN        & P04                             \\ 
23 & 603   &  1.8$\pm$0.1   &   0.54 $\pm$ 0.09  & 00 52 17.51 & -73 01 56.4 & 1.0 & 39.57   & 20.52  & 25.73  & 11.02  & --    & 6.76   & 1.87   & -1.25$\pm$0.08  & AGN        & jet, P04, H00                   \\ 
24 & 645   &  1.0$\pm$0.1   &  -0.06 $\pm$ 0.11  & 01 09 28.34 & -72 22 19.5 & 1.5 & 7.59    & 2.21   & --     & 2.82   & --    & --     & --     & -0.99$\pm$0.10  & galaxy     & jet, P04, S06, J09              \\ 
25 & 647   &  0.7$\pm$0.1   &  -0.03 $\pm$ 0.23  & 01 10 40.59 & -72 28 32.2 & 1.5 & 346.97  & 138.28 & --     & 147.02 & --    & 74.95  & 32.60  & -0.91$\pm$0.05  & AGN        & jet                             \\ 
\noalign{\smallskip}                                                                                                                                                 
26 & 691   &  1.8$\pm$0.2   &   0.57 $\pm$ 0.14  & 00 58 50.27 & -71 36 55.9 & 5.0 & 2.10    & --     & --     & --     & --    & 2.79   & 2.00   & 0.16$\pm$0.20   & AGN        &                                 \\ 
27 & 709   &  0.8$\pm$0.1   &  -0.13 $\pm$ 0.11  & 00 51 17.18 & -73 39 59.6 & 1.0 & 86.00   & 55.24  & 14.38  & 52.28  & --    & 29.83  & 25.16  & -0.33$\pm$0.06  & galaxy     & P04, H00                        \\ 
28 & 746   &  8.6$\pm$0.3   &   0.34 $\pm$ 0.03  & 00 47 19.15 & -72 39 48.1 & 1.0 & 6.96    & 16.27  & 10.61  & 18.15  & --    & 38.72  & 29.94  & 0.66$\pm$0.05   & AGN        & jet, H00                        \\ 
29 & 783   &  2.2$\pm$0.1   &   0.09 $\pm$ 0.05  & 00 57 36.97 & -73 12 58.4 & 1.5 & 3.34    & 2.13   & --     & 2.64   & --    & 2.23   & --     & -0.17$\pm$0.14  & AGN        & H00, P04                        \\ 
30 & 787   &  2.6$\pm$0.3   &   0.07 $\pm$ 0.53  & 00 57 50.22 & -73 05 59.8 & 1.5 & 3.06    & 2.24   & --     & 3.20   & --    & --     & --     & 0.04$\pm$0.25   & AGN        & P04                             \\ 
\noalign{\smallskip}                                                                                                                                                 
31 & 878   &  0.9$\pm$0.1   &   0.19 $\pm$ 0.43  & 01 10 53.21 & -73 14 16.2 & 1.5 & 1244.5  & 575.78 & --     & 555.33 & --    & 273.56 & 139.18 & -0.87$\pm$0.05  & AGN        & jet                             \\ 
32 & 898   &  1.0$\pm$0.2   &   0.51 $\pm$ 0.15  & 00 59 27.28 & -73 08 38.2 & 5.0 & 2.37    & --     & --     & --     & --    & --     & --     & --              & AGN        &                                 \\  
33 & 914   &  4.7$\pm$0.5   &   0.43 $\pm$ 0.12  & 01 16 49.13 & -73 42 38.8 & 1.5 & 5.44    & 2.36   & --     & --     & --    & 2.68   & --     & -0.31$\pm$0.15  & AGN        & P04, H00                        \\ 
34 & 916   &  2.8$\pm$0.3   &   0.45 $\pm$ 0.08  & 01 16 15.86 & -73 26 57.2 & 1.5 & 56.70   & 36.70  & --     & 31.74  & --    & 12.40  & 9.63   & -0.80$\pm$0.06  & ClG        & P04, H00                        \\ 
35 & 946   &  0.4$\pm$0.1   &   0.39 $\pm$ 0.33  & 01 17 09.47 & -73 36 31.7 & 1.5 & 2.97    & 1.29   & --     & --     & --    & --     & --     & -1.64$\pm$0.25  & AGN        &                                 \\ 
\noalign{\smallskip}                                                                                                                                                 
36 & 955   &  0.5$\pm$0.2   &  -0.30 $\pm$ 0.28  & 01 18 33.17 & -73 38 02.5 & 1.5 & --      & --     & --     & --     & --    & --     & 2.07   & --              & AGN        &                                 \\  
37 & 1041  &  752.9$\pm$1.8 &  -0.17 $\pm$ 0.00  & 00 53 07.97 & -74 39 05.2 & 3.0 & 2.40    & 10.79  & --     & 20.90  & --    & 2.96   & 1.43   & -0.47$\pm$0.20  & star       & CF Tuc, P04, H00                \\ 
38 & 1046  &  4.9$\pm$0.3   &   0.08 $\pm$ 0.06  & 00 55 28.33 & -74 44 20.5 & 1.5 & 1.62    & 1.35   & --     & --     & --    & 0.88   & --     & -0.35$\pm$0.20  & AGN        &                                 \\ 
39 & 1047  &  3.1$\pm$0.2   &   0.72 $\pm$ 0.08  & 00 51 54.01 & -74 35 32.2 & 1.5 & --      & 2.24   & --     & --     & --    & 1.16   & --     & -0.53$\pm$0.25  & AGN        & H00                             \\ 
40 & 1145  &  0.3$\pm$0.1   &   0.36 $\pm$ 0.34  & 01 03 36.32 & -72 10 31.6 & 1.0 & --      & --     & 0.43   & --     & --    & --     & --     & --              & AGN        &                                 \\ 
\noalign{\smallskip}                                                                                                                                                 
41 & 1267  &  2.7$\pm$0.3   &   0.01 $\pm$ 0.09  & 00 59 27.59 & -75 12 06.6 & 1.5 & 8.14    & 2.54   & --     & --     & --    & 2.68   & --     & -0.51$\pm$0.20  & AGN        &                                 \\ 
42 & 1490  &  1.4$\pm$0.2   &   0.25 $\pm$ 0.17  & 00 44 38.68 & -73 37 10.5 & 1.0 & 12.20   & 11.03  & 10.48  & 7.37   & --    & 4.84   & --     & -0.57$\pm$0.17  & AGN        & H00                             \\ 
43 & 1533  &  2.5$\pm$0.5   &   0.55 $\pm$ 0.21  & 00 37 57.61 & -72 52 02.6 & 1.5 & 255.86  & 155.12 & --     & 94.44  & --    & 50.97  & 19.82  & -1.06$\pm$0.05  & AGN        & jet                             \\ 
44 & 1540  &  0.8$\pm$0.2   &   0.22 $\pm$ 0.24  & 00 38 57.81 & -72 48 50.8 & 1.5 & 5.70    & 3.50   & --     & --     & --    & 3.37   & --     & -0.25$\pm$0.12  & AGN        & P04                             \\ 
45 & 1595  &  1.1$\pm$0.2   &   0.36 $\pm$ 0.14  & 01 02 37.43 & -72 50 38.5 & 1.5 & --      & 4.12   & --     & --     & --    & 3.48   & --     & -0.14$\pm$0.25  & AGN        &                                 \\ 
\noalign{\smallskip}                                                                                                                                                 
46 & 1607  &  0.8$\pm$0.2   &   0.42 $\pm$ 0.37  & 01 03 39.35 & -72 50 41.4 & 1.5 & 16.39   & 10.29  & --     & 6.73   & --    & 6.72   & 5.60   & -0.42$\pm$0.10  & AGN        & P04                             \\ 
47 & 1688  &  0.5$\pm$0.1   &   0.24 $\pm$ 0.16  & 00 51 41.47 & -72 55 57.7 & 1.0 & 71.88   & 58.23  & 58.34  & 37.53  & --    & 21.29  & 7.26   & -0.97$\pm$0.05  & AGN        & jet, P04                        \\ 
48 & 1726  &  1.2$\pm$0.2   &  -0.01 $\pm$ 0.15  & 01 09 14.47 & -72 29 38.5 & 1.5 & 5.49    & 4.65   & --     & 1.29   & --    & --     & --     & -1.44$\pm$0.12  & galaxy     & J09, S06, P04                   \\ 
49 & 1739  &  0.6$\pm$0.2   &   0.75 $\pm$ 0.45  & 01 10 05.35 & -72 26 47.9 & 1.5 & 115.60  & 132.46 & --     & 135.10 & --    & 72.59  & 41.79  & -0.47$\pm$0.05  & AGN        & jet, P04                        \\ 
50 & 1864  &  1.3$\pm$0.2   &   0.48 $\pm$ 0.14  & 00 49 01.75 & -73 44 55.5 & 1.0 & 12.28   & 6.98   & 2.06   & 6.89   & --    & 5.08   & 0.93   & -0.75$\pm$0.09  & AGN        & P04                             \\ 
\noalign{\smallskip}                                                                                                                                                 
51 & 1870  &  1.1$\pm$0.2   &   0.34 $\pm$ 0.16  & 00 50 11.08 & -73 50 53.0 & 3.0 & --      & --     & --     & --     & --    & --     & 2.16   & --              & AGN        &                                 \\ 
52 & 1910  &  10.4$\pm$0.4  &   0.39 $\pm$ 0.05  & 00 55 18.82 & -71 44 51.0 & 1.5 & 25.92   & 13.40  & --     & 9.74   & --    & 4.57   & --     & -0.97$\pm$0.11  & AGN        & P04                             \\ 
53 & 1936  &  0.9$\pm$0.1   &   0.15 $\pm$ 0.19  & 00 52 06.60 & -71 44 15.1 & 5.0 & 4.54    & --     & --     & --     & --    & 1.90   & --     & -0.50$\pm$0.20  & AGN        &                                 \\ 
54 & 1961  &  0.3$\pm$0.1   &  -0.25 $\pm$ 0.22  & 00 52 55.27 & -71 45 39.1 & 5.0 & 2.74    & --     & --     & --     & --    & --     & --     & --              & AGN/star?  &                                 \\ 
55 & 1990  &  1.0$\pm$0.2   &   0.69 $\pm$ 0.17  & 00 48 25.77 & -72 00 33.7 & 1.5 & --      & 3.16   & --     & --     & --    & 0.92   & --     & -1.00$\pm$0.25  & AGN        &                                 \\ 
\noalign{\smallskip}                                                                                                                                                 
56 & 2055  &  1.3$\pm$0.2   &   0.28 $\pm$ 0.13  & 01 03 31.90 & -71 29 10.8 & 3.0 & --      & --     & --     & 3.36   & --    & --     & --     & --              & AGN        & H00                             \\ 
57 & 2078  &  0.4$\pm$0.1   &  -0.04 $\pm$ 0.20  & 01 04 41.29 & -71 31 23.1 & 1.5 & 19.48   & 24.61  & --     & 26.69  & --    & 18.38  & 10.79  & -0.27$\pm$0.08  & AGN        & P04                             \\ 
58 & 2127  &  0.6$\pm$0.2   &   0.88 $\pm$ 0.25  & 01 02 22.01 & -71 27 20.7 & 1.5 & 13.14   & 11.15  & --     & 10.07  & --    & 4.94   & --     & -0.55$\pm$0.07  & AGN        & P04                             \\ 
59 & 2250  &  0.6$\pm$0.1   &   1.00 $\pm$ 0.26  & 00 49 33.41 & -72 19 01.5 & 1.5 & 23.77   & 20.23  & --     & 18.31  & --    & 7.03   & 4.29   & -0.78$\pm$0.05  & AGN        & jet, P04                        \\ 
60 & 2270  &  0.2$\pm$0.1   &  -0.41 $\pm$ 0.67  & 00 49 48.88 & -72 22 12.8 & 1.5 & 6.15    & 3.01   & --     & --     & --    & --     & --     & -1.41$\pm$0.25  & AGN        &                                 \\ 
\noalign{\smallskip}                                                                                                                                                 
61 & 2302  &  0.4$\pm$0.1   &   0.39 $\pm$ 0.17  & 00 55 57.07 & -72 26 04.7 & 1.5 & 111.95  & 98.53  & --     & 80.32  & --    & 52.37  & 33.53  & -0.52$\pm$0.05  & AGN        & jet                             \\ 
62 & 2368  &  0.3$\pm$0.1   &   0.52 $\pm$ 0.35  & 00 49 36.96 & -72 35 52.4 & 1.5 & --      & 2.25   & --     & 2.36   & --    & --     & 1.54   & -0.23$\pm$0.15  & AGN        &                                 \\ 
63 & 2381  &  5.9$\pm$0.6   &   0.16 $\pm$ 0.09  & 00 42 39.85 & -72 33 24.5 & 1.5 & 12.31   & 10.55  & --     & 8.91   & --    & 3.79   & --     & -0.67$\pm$0.14  & AGN        & P04                             \\ 
64 & 2396  &  0.6$\pm$0.1   &   0.51 $\pm$ 0.18  & 00 44 13.84 & -72 43 00.9 & 1.0 & 18.37   & 15.15  & 3.16   & 19.56  & --    & 8.42   & 2.46   & -0.58$\pm$0.06  & AGN        & P04, K09                        \\ 
65 & 2439  &  0.5$\pm$0.1   &  -0.13 $\pm$ 0.23  & 00 42 26.22 & -73 04 17.8 & 1.0 & 138.90  & 85.61  & 49.56  & --     & --    & 19.95  & 9.65   & -1.09$\pm$0.05  & AGN        & jet, P04                        \\ 
\noalign{\smallskip}                                                                                                                                                 
66 & 2496  &  0.1$\pm$0.1   &   0.46 $\pm$ 0.33  & 00 43 40.76 & -73 25 48.6 & 1.0 & --      & 7.10   & 6.02   & --     & --    & --     & --     & --              & AGN        &                                 \\ 
67 & 2505  &  0.4$\pm$0.1   &  -0.14 $\pm$ 0.32  & 00 42 15.25 & -73 29 08.8 & 2.0 & --      & 1.69   & --     & --     & --    & --     & --     & --              & AGN        &                                 \\ 
68 & 2556  &  0.3$\pm$0.1   &  -0.40 $\pm$ 0.73  & 00 48 58.33 & -73 24 39.5 & 1.0 & --      & --     & 0.22   & --     & --    & --     & --     & --              & AGN        &                                 \\ 
69 & 2568  &  1.0$\pm$0.2   &   0.74 $\pm$ 0.17  & 00 52 38.31 & -73 12 45.0 & 1.0 & 143.94  & 98.95  & 105.37 & 72.18  & --    & 33.25  & 15.85  & -0.95$\pm$0.05  & AGN        & jet, P04                        \\ 
70 & 2575  &  0.6$\pm$0.1   &   0.76 $\pm$ 0.21  & 00 50 57.54 & -73 12 48.0 & 1.0 & 2.06    & 2.93   & 4.08   & --     & 4.10  & 3.57   & 2.47   & -0.01$\pm$0.09  & AGN        &                                 \\ 
\noalign{\smallskip}                                                                                                                                                 
71 & 2609  &  0.4$\pm$0.1   &   0.55 $\pm$ 0.19  & 00 55 54.47 & -73 03 43.9 & 2.0 & 4.74    & 2.46   & --     & --     & --    & --     & --     & -1.29$\pm$0.25  & AGN        &                                 \\ 
72 & 2695  &  3.2$\pm$0.4   &   0.38 $\pm$ 0.11  & 01 02 16.60 & -72 37 04.6 & 1.0 & --      & --     & 3.26   & --     & --    & --     & --     & --              & ClG        & jet                             \\ 
73 & 2712  &  0.8$\pm$0.2   &   1.00 $\pm$ 0.15  & 01 05 38.47 & -72 23 02.1 & 1.0 & --      & --     & 1.39   & --     & 1.09  & 2.35   & --     & 0.43$\pm$0.30   & AGN        & P04                             \\ 
74 & 2738  &  4.9$\pm$0.3   &   0.41 $\pm$ 0.07  & 01 10 50.08 & -72 10 26.7 & 1.5 & 16.15   & 5.63   & --     & 7.52   & --    & 8.25   & 4.03   & -0.39$\pm$0.08  & AGN        & jet, P04                        \\ 
75 & 2776  &  0.5$\pm$0.1   &  -0.50 $\pm$ 0.39  & 01 10 00.34 & -72 08 24.5 & 1.5 & 3.16    & 2.53   & --     & --     & --    & 1.42   & 1.18   & -0.43$\pm$0.09  & AGN        &                                 \\ 
\noalign{\smallskip}                                                                                                                                                 
76 & 2795  &  2.0$\pm$0.2   &   0.42 $\pm$ 0.08  & 01 12 44.48 & -72 24 21.3 & 3.0 & --      & --     & --     & --     & --    & --     & 3.29   & --              & AGN        &                                 \\ 
77 & 2834  &  0.4$\pm$0.1   &   0.86 $\pm$ 0.21  & 01 14 13.77 & -72 18 19.0 & 3.0 & --      & --     & --     & --     & --    & 2.32   & 1.78   & -0.45$\pm$0.25  & AGN        &                                 \\ 
78 & 2853  &  0.5$\pm$0.1   &   0.43 $\pm$ 0.16  & 01 11 19.53 & -72 49 01.9 & 3.5 & --      & --     & --     & --     & --    & --     & 2.59   & --              & AGN        & jet                             \\ 
79 & 2905  &  6.3$\pm$0.4   &   0.19 $\pm$ 0.06  & 01 11 32.47 & -73 02 09.6 & 1.5 & 69.22   & 70.02  & --     & 74.88  & --    & 63.25  & 62.74  & -0.05$\pm$0.05  & galaxy     & jet, J09, S06, P04, H00         \\ 
80 & 2978  &  0.2$\pm$0.1   &   0.42 $\pm$ 0.41  & 01 15 26.14 & -73 00 18.3 & 2.0 & 2.64    & 2.35   & --     & --     & --    & --     & --     & -0.23$\pm$0.25  & AGN        &                                 \\ 
\noalign{\smallskip}                                                                                                                                                 
81 & 2987  &  3.7$\pm$0.3   &   0.39 $\pm$ 0.07  & 01 21 07.26 & -73 07 06.8 & 2.0 & 2.20    & 2.10   & --     & --     & --    & --     & --     & -0.09$\pm$0.25  & AGN        & jet, H00                        \\ 
82 & 3053  &  1.3$\pm$0.1   &   0.31 $\pm$ 0.10  & 01 13 22.87 & -73 24 53.5 & 3.5 & --      & --     & --     & --     & --    & --     & 3.11   & --              & AGN        & K09                             \\ 
83 & 3117  &  1.1$\pm$0.1   &   0.35 $\pm$ 0.09  & 01 06 58.03 & -73 13 20.9 & 2.0 & 7.22    & 2.65   & --     & --     & --    & 0.93   & --     & -1.12$\pm$0.20  & AGN        &                                 \\ 
84 & 3150  &  0.7$\pm$0.2   &  -0.26 $\pm$ 0.19  & 01 10 21.91 & -73 04 39.4 & 1.5 & 70.11   & 48.71  & --     & 37.29  & --    & 25.47  & 12.81  & -0.69$\pm$0.05  & AGN        & jet, P04                        \\ 
85 & 3163  &  0.3$\pm$0.1   &   0.95 $\pm$ 0.35  & 01 07 08.37 & -73 07 19.7 & 2.0 & 1.96    & 1.49   & --     & 1.51   & --    & --     & --     & -0.26$\pm$0.16  & AGN        &                                 \\ 
\noalign{\smallskip}                                                                                                                                                 
86 & 3177  &  0.5$\pm$0.2   &  -0.63 $\pm$ 0.40  & 01 00 09.87 & -73 13 12.4 & 2.0 & 2.79    & 1.73   & --     & --     & --    & --     & --     & -0.94$\pm$0.25  & galaxy     & S06                             \\ 
87 & 3208  &  0.9$\pm$0.2   &   0.07 $\pm$ 0.32  & 00 58 15.56 & -73 16 14.1 & 5.0 & 4.66    & --     & --     & --     & --    & --     & --     & --              & galaxy     & J09, S06                        \\ 
88 & 3209  &  0.1$\pm$0.1   &  -0.03 $\pm$ 0.27  & 00 57 49.36 & -73 25 44.9 & 5.0 & 2.62    & --     & --     & --     & --    & --     & --     & --              & AGN        &                                 \\

\end{longtable}

\tablefoot{For a description of the Table, see Sec.~\ref{sec:data:correlation}.
\tablefoottext{a}{Comments on individual sources. For details see text.
Identifications with sources from other catalogues are marked as follows:}
}
\tablebib{
(P04) \citet{2004MNRAS.355...44P};
(H00) \citet{2000A&AS..142...41H};
(J09) \citet{2009MNRAS.399..683J};
(S06) \citet{2006AJ....131.1163S};
(S05) \citet{2005MNRAS.362..879S};
(K09) \citet{2009ApJ...701..508K}.
}
\end{landscape}
}

\end{document}